\documentclass[10pt,journal,compsoc]{IEEEtran}

\ifCLASSOPTIONcompsoc
  \usepackage[nocompress]{cite}
\else
  \usepackage{cite}
\fi

\usepackage{tikz}
\usepackage{amsmath}
\usepackage{amssymb,amsfonts}
\usepackage{tabularx}
\usepackage{booktabs} 
\usepackage[edges]{forest}
\usepackage{graphicx}
\usepackage{url}

\begin{document}

\thispagestyle{empty}
\onecolumn

{\Large \bfseries Gotham Testbed: a Reproducible IoT Testbed for Security Experiments and Dataset Generation}

\vspace{3cm}

{\LARGE This article has been accepted for publication in IEEE Transactions on Dependable and Secure Computing. This is the author's version which has not been fully edited and content may change prior to final publication.}

\vspace{1cm}

{\LARGE Citation information: \url{https://doi.org/10.1109/TDSC.2023.3247166}}

\twocolumn
\clearpage
\pagenumbering{arabic}

\title{Gotham Testbed: a Reproducible IoT Testbed for Security Experiments and Dataset Generation}

\author{Xabier~Sáez-de-Cámara,
        Jose~Luis~Flores,
        Cristóbal~Arellano,
        Aitor~Urbieta,
        and~Urko~Zurutuza%
\IEEEcompsocitemizethanks{\IEEEcompsocthanksitem X. Sáez-de-Cámara, J.L. Flores, C. Arellano and A. Urbieta are with Ikerlan Technology Research Centre, Basque Research and Technology Alliance (BRTA), Arrasate-Mondragón, Spain.\\E-mail: \{xsaezdecamara, jlflores, carellano, aurbieta\}@ikerlan.es
\IEEEcompsocthanksitem U. Zurutuza and X. Sáez-de-Cámara are with Mondragon Unibertsitatea, Arrasate-Mondragón, Spain. E-mail: uzurutuza@mondragon.edu, xabier.saezdecamara@alumni.mondragon.edu}}

\IEEEtitleabstractindextext{%
\begin{abstract}
	The growing adoption of the Internet of Things (IoT) has brought a significant increase in attacks targeting those devices. Machine learning (ML) methods have shown promising results for intrusion detection; however, the scarcity of IoT datasets remains a limiting factor in developing ML-based security systems for IoT scenarios. Static datasets get outdated due to evolving IoT architectures and threat landscape; meanwhile, the testbeds used to generate them are rarely published. This paper presents the Gotham testbed, a reproducible and flexible security testbed extendable to accommodate new emulated devices, services or attackers. Gotham is used to build an IoT scenario composed of 100 emulated devices communicating via MQTT, CoAP and RTSP protocols, among others, in a topology composed of 30 switches and 10 routers. The scenario presents three threat actors, including the entire Mirai botnet lifecycle and additional red-teaming tools performing DoS, scanning, and attacks targeting IoT protocols. The testbed has many purposes, including a cyber range, testing security solutions, and capturing network and application data to generate datasets. We hope that researchers can leverage and adapt Gotham to include other devices, state-of-the-art attacks and topologies to share scenarios and datasets that reflect the current IoT settings and threat landscape.
\end{abstract}

}

\maketitle

\IEEEdisplaynontitleabstractindextext

\IEEEpeerreviewmaketitle

\IEEEraisesectionheading{\section{Introduction}\label{sec:introduction}}

\IEEEPARstart{T}{he} Internet of Things (IoT) and Machine to Machine (M2M) communication protocols are rapidly developing technologies of great interest to the industrial sector. They have the potential to improve the efficiency and reliability of manufacturing operations and processes, as well as foster the creation of new products, applications and business models~\cite{Wortmann2015}. Due to the pervasiveness of IoT, security and privacy guarantees should be one of the main concerns to be addressed. Unfortunately, multiple sources of vulnerabilities such as deficient physical security, inadequate authentication, improper encryption, unnecessary open ports, insufficient access control, improper patch management, weak programming practices and insufficient audit mechanisms are the main reasons why many of these types of devices are currently susceptible to attacks~\cite{Neshenko2019}. This situation has led to the development of malware specifically designed to target and exploit IoT devices~\cite{Antonakakis2017, Vervier2018}, typically incorporating the compromised machines into a botnet to launch multiple campaigns such as Distributed Denial of Service (DDoS) attacks, spamming, cryptocurrency mining or advertisement click fraud~\cite{Kambourakis2019}. Exposed Industrial IoT (IIoT) systems, which are often part of critical infrastructures, are also the targets of many attacks, including ransomware~\cite{DragosRansomware}, intellectual property theft or sabotage~\cite{Sadeghi2015}.

To defend against these types of cybersecurity threats, in addition to the use of traditional rule-based intrusion detection and prevention systems (IDS/IPS), there has been a vast body of research on statistical modeling, machine learning (ML) and deep learning (DL) methods over the last three decades. However, despite the promising results of ML in cybersecurity~\cite{Ferrag2020}, the deployment of these systems is still limited in practice, creating a gap between the academic settings and operational environments~\cite{Sommer2010, Arp2022}.

One of the main reasons for this gap is the lack of representative public datasets that include up-to-date traffic and attacking patterns to develop ML-based systems~\cite{Sommer2010, Ferrag2020, Berman2019}. This issue is particularly relevant for IoT and M2M environments, where the number of special-purpose public datasets is currently insufficient~\cite{Hindy2020}, and the attacking behaviors included in the datasets are outdated or underrepresented due to the rapidly evolving IoT threat landscape~\cite{Vervier2018, costin2018iot}.

Using real IoT hardware on operational networks is one of the preferred methods to generate an accurate and representative dataset. Nevertheless, testing on real networks is not always feasible. Network traffic data can include confidential and personally identifiable information, making it difficult to publish. Experimentation in real deployments can also be challenging, time-consuming and expensive. Furthermore, using real malware samples or attacking tools to generate representative threats can potentially harm the devices and raise ethical considerations~\cite{Sommer2010, Siaterlis2013, ElSheikh2014, Arp2022}.

An alternative to real settings is the use of emulation-based systems~\cite{Siaterlis2013}. Emulation software enables researchers to create network testbeds composed of multiple emulated devices that can be used for various purposes, including evaluating security solutions, testing network topologies, personnel training exercises or other research. The activities performed within the testbed generate traces (such as network packets, system logs, and syscalls) that can be captured and used, for instance, to create datasets to develop or evaluate ML models for the detection of attacks or malicious behavior performed in the testbed. While many of the datasets created from emulated testbeds are openly available to the community, the testbeds themselves are rarely published. This prevents other researchers from adapting the testbed, and the generated data, to their particular use cases. Moreover, there is a generalized lack of documentation about the configuration used in the testbeds to generate the datasets~\cite{Hindy2020}. The scarcity of details on the devices, software versions, and configuration files or command-line arguments used in attacking tools hinders the reproducibility of the datasets. This issue was recently raised in~\cite{Engelen2021} when analyzing the popular CICIDS2017~\cite{Sharafaldin2018} dataset.

We argue that sharing datasets alone for ML model training is not enough to reduce the gap between the experimental and deployment environments. Especially considering that in some cases, ML models learned from public datasets may not generalize well to other network settings~\cite{Catillo2022}. Sharing a reproducible and extendable testbed allows researchers and practitioners to leverage and adapt the platform to be as close as possible to the network setting of interest. To that end, we present a testbed for IoT security research that allows performing security experiments and the extraction of real network traffic datasets. The contributions can be summarized as follows:

\begin{itemize}
	\item An IoT network security testbed implemented as a middleware over the GNS3 network emulator~\cite{Grossmann}. It allows the deployment of different network topologies and is flexible enough to incorporate any type of physical, virtualized or containerized clients, servers and applications, as well as generate real network traffic data. The source code to reproduce the testbed is available at~\cite{MiRepositorio}.

	\item A ready-to-use scenario composed of 100 emulated IoT and IIoT devices, servers and attackers. The devices are connected in a realistic topology with 30 network switches and 10 routers. Particularly, the emulated IoT nodes communicate primarily via the MQTT and CoAP M2M protocols and the RTSP streaming protocol.

	\item A threat model that includes 3 different threat actors executing real botnet malware and other red-teaming attack tools. The testbed includes the (i) Mirai worm~\cite{Antonakakis2017} and all its required command and control (C\&C) infrastructure; (ii) a second botnet based on the Merlin C\&C server~\cite{Tuyl} and (iii) network scans and attacks specifically targeting the MQTT and CoAP services.

	\item We provide a set of properties based on the literature that an emulation platform should meet, and validate the proposed platform based on those properties.
\end{itemize}

The rest of this paper is structured as follows. Section~\ref{sec:relatedwork} discusses the related work. Section~\ref{sec:testbedproperties} details the general properties to be met by network security testbeds. The testbed architecture is presented in Section~\ref{sec:testbedarch}, and the IoT use-case scenario is detailed in Section~\ref{sec:testbedimplementation}. Section~\ref{sec:evaluation} shows experimental results related to the properties of the testbed. Finally, the paper is concluded in Section~\ref{sec:discussion}.

\section{Related Work} \label{sec:relatedwork}

In this section, several publications about IoT testbeds and datasets for network security are described and discussed.

\subsection{Testbeds and datasets for IoT security}

Many IoT datasets are usually generated using a testbed composed of real or emulated devices. Meidan et al.~\cite{Meidan2018} present N-BaIoT, a dataset generated using a small laboratory setup composed of 9 real commercial IoT devices. They deploy Mirai and BASHLITE botnets to capture traffic in both normal and compromised states. However, they do not consider the whole botnet lifecycle and only focus on the DoS attacking stages; the botnet propagation, infection and communication with the command and control server stages are not included. The deployment does not represent a realistic network topology because all the IoT devices and servers needed for the botnet infrastructure are located in the same LAN connected to a single switch. Raw network traces in pcap format are not available; only processed features are included.

Koroniotis et al.~\cite{Koroniotis2019} design a testbed composed of emulated IoT devices as well as emulated PCs and servers, which is used to extract the Bot-IoT network traffic dataset. They include a total of 8 Windows and Linux virtual machines (VM) to implement the normal and attacking nodes, all of them connected to the same LAN. Node-red is used to emulate the traffic of 5 IoT sensors that send messages via the MQTT protocol to a public AWS broker in addition to the Ostinato traffic generator to simulate normal network activity. They use several Kali Linux VMs to perform the attacks. As an evolution from the previous work, Moustafa~\cite{Moustafa2021} builds an emulated testbed architecture including a mix of IoT devices and regular IT clients to generate the TON\_IoT dataset. The dataset includes network traffic, application and OS logs. The testbed is composed of 17 VMs and includes a single VM simulating 7 different IoT sensors, a smart TV, 2 smartphones and several client systems based on Windows, Linux and purposefully vulnerable VMs. As in the previous version, they use Node-red to generate the MQTT IoT traffic to a public broker and the Ostinato traffic generator for the rest of the normal traces. The previous datasets lack attack heterogeneity, real botnet malware is not included, and while they contain a diverse set of attacks, they do not include attacks targeted against the MQTT IoT protocol.

Hindy et al.~\cite{Hindy2021} publish the MQTT-IoT-IDS2020 dataset to evaluate the effectiveness of ML techniques to detect MQTT-based attacks. The dataset is generated using a VM-based emulated testbed composed of 12 simulated MQTT sensors publishing random messages of varying length to a single broker, two machines simulating a UDP stream and one attacker. All the sensors are located at the same LAN, while the broker, stream server and the attacker are in another network separated by a single router. The attacks are limited to generic network scans and an MQTT brute force attack.

IoT-Flock~\cite{Ghazanfar2020} is a traffic generator to simulate MQTT and CoAP-based IoT devices and attacks. Vaccari et al.~\cite{Vaccari2020} present MQTTset; they use IoT-Flock to simulate 8 simple sensors publishing data to an MQTT broker and a single malicious node that can launch DoS attacks, malformed data and brute force attacks. All the emulated devices are directly connected to the same LAN. The deployment does not represent a realistic network topology, and the attacks only target the broker. Similarly, Hussain et al.~\cite{Hussain2021} use IoT-Flock to create a dataset consisting of MQTT and CoAP network traces. They create a testbed mimicking an IoT-based healthcare system with 9 emulated simple sensors, an MQTT broker and a CoAP server. The attacks include MQTT packet floods, packet crafting and CoAP replay attacks.

Guerra-Manzanares et al.~\cite{GuerraManzanares2020} design a testbed composed of real and emulated devices to generate the MedBioT dataset. They use 3 real commercial IoT devices and 4 emulated MQTT-based IoT sensor templates, of which they instantiate 20 of each using containerization technologies for a total of 83 devices. Three real botnet malware samples (Mirai, BASHLITE and Torii) are included to generate the attacks. The larger scale of this testbed enables a more realistic botnet propagation pattern compared to smaller ones. However, all the IoT devices and the botnet infrastructure are directly connected to a single switch in the same LAN, which does not reflect a realistic IoT topology. Additionally, they only focus on the first stages of the botnet lifecycle (infection, propagation and command and control communication), but they neither include attacking stages nor IoT-specific attacks. They do not provide details about the source code modifications and command and control infrastructure configuration needed to run real malware in the testbed.

Amine Ferrag et al.~\cite{Ferrag2022} present Edge-IIoTset, an IoT/IIoT dataset that includes MQTT and Modbus traffic generated using a testbed composed of real low-cost sensors and emulated devices. For the IoT and IIoT devices, they wire 11 sensors to an Arduino Uno board; they deploy MQTT brokers and Modbus master/slave nodes using the Node-red Modbus extension on various Raspberry Pi boards. They also emulate multiple vulnerable services, applications and attackers using VMs. There are no precise details about the total number of nodes, configuration options and network topology; however, all the attackers and victims seem to be connected to the same wireless router, which does not reflect a realistic IoT topology. They perform a varied set of attacks, yet, most attacks target the services in the vulnerable VMs instead of the IoT devices. The attacks targeting the IoT nodes include generic flooding, scanning and spoofing attacks, but attacks against IoT protocols are lacking.

The literature presents some works that share some similarities to our proposed testbed. Antonioli et al.~\cite{Antonioli2015} present MiniCPS, an extensible and reproducible testbed to emulate communication in CPSs such as PLCs and HMIs using the Ethernet/IP and Modbus TCP/IP protocols. They use Mininet for the network emulation layer. The purpose of the testbed is to perform attacks on CPS systems and develop defenses; they provide an example of ARP spoofing and man-in-the-middle (MITM) traffic manipulation attacks and develop a detection method using a custom software-defined network controller. Eckhart et al.~\cite{Eckhart2018} present CPS Twinning, a Mininet-based testbed for creating digital twins used to test or monitor security/safety rules and data capturing purposes. CPS Twinning includes a generator module that can automatically create the virtual testbed in a reproducible way based on parsing specification files defined in the AutomationML data format. The prototype includes PLCs and HMIs running native code and communicating with Modbus TCP/IP protocol. The threat scenario presents an ARP spoofing and MITM attack and shows successful detection by monitoring various states of the digital twin.

Our testbed differs from both cited testbeds in~\cite{Antonioli2015} and~\cite{Eckhart2018} in several ways. In~\cite{Antonioli2015}, creating a new type of emulated host or adding support for another communication protocol requires modifications to the testbed code and porting the code to Python, limiting its extensibility to add heterogeneous nodes communicating with diverse protocols. CPS Twinning~\cite{Eckhart2018} assumes that the organization already uses the AutomationML language to define its physical infrastructure and is only focused on Modbus TCP/IP. Each emulated host in our approach is intended to run arbitrary programs and communicate using arbitrary protocols over TCP/IP. The created scenarios in both testbeds lack attack diversity and do not include real malware samples. In contrast, we provide an extensive threat model that includes real malware samples to generate various attacks. From the implementation point of view, we use GNS3 to manage the network layer, and we use Docker-based containers (and VMs) to provide a reproducible specification and emulation for each host. Mininet-based testbeds use a lighter containerization model where each host is a group of processes in a network namespace, but all share the same filesystem by default. However, due to the use of real malware samples, we use Docker-based virtualization for a more comprehensive isolation at the expense of greater virtualization overhead. Additionally, while Mininet can impose CPU resource constraints in the emulated hosts, currently, memory constraints are not supported, which limits the fidelity to emulate each host's hardware resources compared to Docker-based hosts.


\subsection{General IoT simulators and testbeds}

Multiple IoT simulators and testbeds are currently available for general IoT network research and development purposes~\cite{Chernyshev2018}. Most of those simulators, such as ns-3, OMNeT++ and CupCarbon, are discrete-event simulators specialized in the physical and media access layer protocol simulations. However, those simulators still lack the support for many application layer IoT protocols out of the box, and are not specifically designed for cybersecurity applications~\cite{Chernyshev2018}. Some recent versions of those simulators support protocols such as MQTT~\cite{CupCarbon}, but the integration of arbitrary application protocols is still lacking. In contrast, in this work, we are interested in emulating devices, servers and network equipment that run real production libraries, network switching software and routing operating systems, as well as real malware samples and attack tools.

\subsection{Discussion}

In general, the presented cybersecurity datasets lack heterogeneity in terms of attacks; most do not include real botnet malware samples, one of the most prominent threats to current IoT devices~\cite{Vervier2018}, and the ones that do include them~\cite{Meidan2018, GuerraManzanares2020} are limited to some botnet stages instead of the whole lifecycle. Additionally, only a few include attacks targeting IoT protocols such as MQTT and CoAP. Furthermore, the scalability of the testbeds is small (in the order of 10 devices, usually less than 20) due to the use of real devices or VMs, except for the work at~\cite{GuerraManzanares2020}, which uses containerization technology. Moreover, most testbeds represent only simplified topologies where all the devices and attackers are connected to the same LAN, which can lead to unrealistic threat models. Testbeds need to include multiple networks and routing layers to represent a realistic botnet propagation and attacking scenario.

There is also a general lack of documentation regarding IoT node behavior, server configuration options (e.g., MQTT broker configuration) and the exact implementation or parameters used to perform the attacks. This information is crucial because many attacks can behave differently depending on the configuration of both the victim and attacker. For example, some MQTT broker implementations and versions are not affected by the MQTT authentication bypass and packet crafting attacks included in~\cite{Hussain2021}, rendering those attacks irrelevant. Similarly, real botnet source code needs certain modifications to make them work in a testbed or limit some potential threats to external networks; however, those patches are not provided in the datasets that include real botnets.

To the best of our knowledge, the cybersecurity testbeds used to generate the cited datasets are not published, except for the testbeds in~\cite{Antonioli2015} and~\cite{Eckhart2018}. This limits extendibility and reproducibility because it prevents other researchers from building upon, reusing or adapting the testbed to generate specialized datasets that best suit their needs.

\section{Testbed requirements and platform features} \label{sec:testbedproperties}

This section details the main requirements that have been defined in the literature for the creation and evaluation of testbeds and datasets to provide a rigorous experimentation platform. We classify and group those requirements to provide a list of the features to be fulfilled by the Gotham testbed.

\subsection{General testbed and dataset requirements}

Over the last years, the community has defined a set of requirements for network testbeds and datasets that should be met to provide accurate and reliable results. According to Siaterlis et al.~\cite{Siaterlis2013, Siaterlis2013a}, the basic testbed requirements include fidelity to reproduce a real system to the sufficient level of detail needed for the current experiment, a controlled environment to allow the reproducibility of the scenarios, and being able to correctly measure and monitor the experiment. Additionally, the testbeds should be comprised of heterogeneous elements, be extendable to include new protocols or devices and be scalable to support networks with many nodes~\cite{Chernyshev2018, Lai2021}.

For testbeds designed to run security experiments, additional requirements have been defined given the presence of malware and attack tools. These requirements include the safe execution of malicious software without interfering with the testbed~\cite{Siaterlis2013a}, containment to prevent the transmission of attacks to an external operational network~\cite{Huang2012} and the ability to emulate scenarios with complex topologies to properly study the whole botnet lifecycle~\cite{ElSheikh2014}.

Many of the defined requirements for testbeds also overlap with the criteria that network security datasets should meet~\cite{Sharafaldin2018, Engelen2021, Hindy2020}. The criteria can be summarized as follows: the dataset provides real and complete network traces; the traffic is generated using a valid network topology that includes clients, servers and network equipment; the dataset is labeled to distinguish between benign and malicious traces; highly heterogeneous regarding included services, network protocols, normal and attack behaviors; easily extendable; reproducible; shareable and documented.

\subsection{Required testbed features}

Considering the requirements from the literature summarized in the previous subsection, we classify and group them into five main properties: fidelity, heterogeneity, scalability, reproducibility and measurability.

Broadly, \textit{fidelity} refers to the ability to reproduce the hardware and software of all the components to a sufficient level of detail and being able to do it without the need for external resources for increased isolation when dealing with malware. \textit{Heterogeneity} refers to the diversity of behaviors, especially if the data is used for ML model training. \textit{Scalability} refers to the ability to create networks with a large number of nodes. \textit{Reproducibility} is needed to enable replication of the results and building upon them to keep improving the platform. \textit{Measurability} refers to the ability and easiness of extracting relevant data from the testbed.

For each property, we derive a set of desired features that should be met to create a comprehensive testbed platform, as shown in \figurename~\ref{fig:testbed_properties}. In the following, we describe the rationale for each feature:

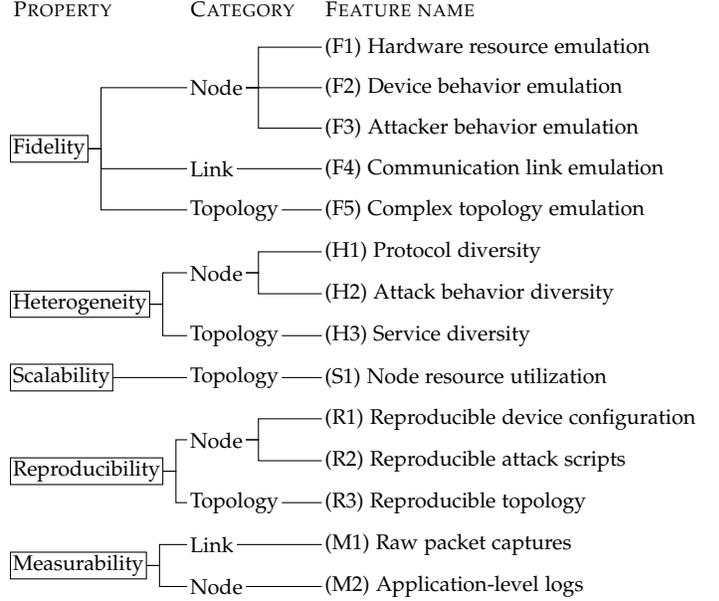
\begin{figure}[t]
	\centering
	\footnotesize
	\begin{forest}
		for tree={grow=east, child anchor=west, anchor=west, inner sep=1pt}, 
		forked edges,
		[,phantom
		[Requirements,phantom
			[Measurability, draw, tier=prop
				[Node, tier=cat
					[(M2) Application-level logs, tier=name]
				]
				[Link, tier=cat
					[(M1) Raw packet captures, tier=name]
				]
			]
			[Reproducibility, draw, tier=prop
				[Topology, tier=cat
					[(R3) Reproducible topology, tier=name]
				]
				[Node, tier=cat
					[(R2) Reproducible attack scripts, tier=name]
					[(R1) Reproducible device configuration, tier=name]
				]
			]
			[Scalability, draw, tier=prop
				[Topology, tier=cat
					[(S1) Node resource utilization, tier=name]
				]
			]
			[Heterogeneity, draw, tier=prop
				[Topology, tier=cat
					[(H3) Service diversity, tier=name]
				]
				[Node, tier=cat
					[(H2) Attack behavior diversity, tier=name]
					[(H1) Protocol diversity, tier=name]
				]
			]
			[Fidelity, draw, tier=prop
				[Topology, tier=cat
					[(F5) Complex topology emulation, tier=name]
				]
				[Link, tier=cat
					[(F4) Communication link emulation, tier=name]
				]
				[Node, tier=cat
					[(F3) Attacker behavior emulation, tier=name]
					[(F2) Device behavior emulation, tier=name]
					[(F1) Hardware resource emulation, tier=name]
				]
			]
		]
		[One,phantom
			[\textsc{Property},no edge,tier=prop
				[\textsc{Category},no edge,tier=cat
					[\textsc{Feature name},no edge,tier=name]]]]
		]
	\end{forest}
	\caption{Required testbed features grouped under different properties and categories.}
	\label{fig:testbed_properties}
\end{figure}

\subsubsection{Fidelity}
This property is further divided in terms of three different categories describing the basic elements of a networked system: nodes (devices, servers, switches, routers, etc.), links (network links between the nodes) and network topology (the arrangement of nodes and links).

\textbf{F1}: Fidelity in terms of node hardware resource emulation. To allow the emulation of devices with different computational capabilities, the testbed should be able to adjust the memory and CPU resources assigned to each node.

\textbf{F2}: Fidelity in terms of node behavior emulation. All IoT nodes, as well as servers, switches and routers, should support running real production applications, libraries and operating systems to generate real network traffic and logs.

\textbf{F3}: Fidelity in terms of attacker behavior emulation. The testbed should support and provide nodes running real malware samples found in the wild and popular red teaming tools.

\textbf{F4}: Fidelity in terms of communication link emulation. Many nodes, especially IoT devices, can be connected to the internet using links of different quality. The testbed should allow the modification of network link QoS properties such as bandwidth limits, delay, jitter and packet loss to emulate different network link types.

\textbf{F5}: Fidelity to emulate complex topologies. The ability to represent real-world network deployments with many clients, servers, switches and routers is necessary to correctly represent several attacks, including botnet propagation, network-wide scans and DDoS attacks. The testbed should also provide all the necessary services and infrastructure (command and control, name resolution, databases, etc.) for the actual malware samples in a contained and isolated manner to avoid leaking traffic or attacks into the Internet.

\subsubsection{Heterogeneity}
This property is described in terms of node and topology category levels.

\textbf{H1}: Heterogeneity in terms of node protocols. The testbed should include nodes that communicate using a diverse set of network protocols. The diversity includes IoT nodes sending telemetry using different protocols, routers communicating with each other topology information using routing protocols and network services such as name resolution.

\textbf{H2}: Heterogeneity in terms of attacks. Besides offering different types of attacks, diversity within the same attack type should also be provided. This procedure includes combining different attack tools that perform similar actions and using multiple options and flags for each attack~\cite{Engelen2021}.

\textbf{H3}: Heterogeneity in terms of services in the topology. Devices and attackers behave differently depending on the configuration of the service; hence protocol heterogeneity (\textbf{H1}) might not be enough for a realistic emulation. The testbed should also provide multiple equivalent services configured in different ways. For instance, services with or without authentication, communicating in plain text or over an encrypted channel.

\subsubsection{Scalability}
It is defined at the topology level.

\textbf{S1}: Scalability to support topologies with many nodes. IoT networks are usually large scale; the ability to include many nodes can increase the realism of the emulated network.

\subsubsection{Reproducibility}
Defined at the node and topology level, the following features should be included and documented to enable a reproducible scenario.

\textbf{R1}: Reproducibility in terms of node configuration. Description of the behavior of each node, including all the programs executing in the node and their configuration.

\textbf{R2}: Reproducibility in terms of attack scripts. Description of the performed attacks, including software, configuration and command-line options.

\textbf{R3}: Reproducibility in terms of topology description. The way in which all the nodes are connected to form the network topology, including the network link properties, should be detailed.

\subsubsection{Measurability}
Defined in terms of link and node categories.

\textbf{M1}: Ability to measure raw network packets from any node. Different experiments might need to capture network traffic at many locations. The testbed should provide packet capturing from arbitrary links in the topology.

\textbf{M2}: Ability to measure application-level logs. Some security solutions work with application-level logs; the testbed should provide this type of data to create datasets of heterogeneous sources. Other additional measurements could include node CPU and memory resource usage metrics.

With the ability to emulate complex topologies (\textbf{F5}) that include all the necessary services and devices, the platform can be entirely isolated from the network and thus prevent the possible propagation of attacks from the testbed to the outside network. Extensibility is also achieved thanks to the reproducibility of all the nodes (\textbf{R1}) (\textbf{R2}). The reproducibility allows other researchers to adapt existing nodes or create new ones to suit their needs. In addition, (\textbf{R2}) also allows labeling of the datasets generated by the testbed.


\subsection{Comparison with related work}

\begin{table*}
	\small
	\centering
	\caption{Comparison With Related Work Based on the Required Testbed Features (\figurename~\ref{fig:testbed_properties})}
	\label{tab:compare_sota_feats}
	\newcommand{\xSI}{\checkmark}
	\newcommand{\xNO}{x} 
	\newcommand{\xNA}{-}
	\newcommand{\xPA}{$\sim$} 
	\newcolumntype{C}{@{\hspace{2pt}}c@{\hspace{2pt}}}
	\begin{tabularx}{1.0\linewidth}{l|X| @{\hspace{6pt}}c@{\hspace{2pt}} C C C C C C C C C C C C @{\hspace{2pt}}c@{\hspace{6pt}} | p{1.35cm} | p{1.35cm}}
		\toprule
		\textbf{Reference}                           & \textbf{Type}            & \textbf{F1}   & \textbf{F2}   & \textbf{F3}   & \textbf{F4}   & \textbf{F5}   & \textbf{H1}   & \textbf{H2}   & \textbf{H3}   & \textbf{S1}   & \textbf{R1}   & \textbf{R2}   & \textbf{R3}   & \textbf{M1}   & \textbf{M2}   & \textbf{Testbed available} & \textbf{physical layer IoT protocols} \\
		\midrule
		N-BaIoT~\cite{Meidan2018}           & real            & \xSI & \xSI & \xSI & \xNA & \xNO & \xNA & \xNO & \xNA & \xNO & \xNO & \xNO & \xNO & \xPA & \xNO & \xNO  & \xNO    \\
		Bot-IoT~\cite{Koroniotis2019}       & VM              & \xNA & \xSI & \xNO & \xNA & \xNO & \xNO & \xPA & \xNA & \xNO & \xNO & \xSI & \xNO & \xPA & \xNO & \xNO  & \xNO    \\
		TON\_IoT~\cite{Moustafa2021}        & VM              & \xNA & \xSI & \xNO & \xNA & \xNO & \xNO & \xSI & \xNA & \xNO & \xNO & \xSI & \xNO & \xPA & \xSI & \xNO  & \xNO    \\
		Hindy~\cite{Hindy2021}              & VM              & \xNA & \xSI & \xNO & \xPA & \xNO & \xNO & \xNO & \xNO & \xNO & \xNO & \xNO & \xNO & \xPA & \xNO & \xNO  & \xNO    \\
		MQTTset~\cite{Vaccari2020}          & IoT-Flock       & \xNA & \xSI & \xNO & \xNA & \xNO & \xNO & \xNO & \xNO & \xNO & \xNO & \xSI & \xNO & \xPA & \xNO & \xNO  & \xNO    \\
		Hussain~\cite{Hussain2021}          & IoT-Flock       & \xNA & \xSI & \xNO & \xNA & \xNO & \xNO & \xNO & \xNO & \xNO & \xNO & \xSI & \xNO & \xPA & \xNO & \xNO  & \xNO    \\
		MedBIoT~\cite{GuerraManzanares2020} & real, container & \xSI & \xSI & \xSI & \xNO & \xNO & \xNA & \xNO & \xNO & \xSI & \xNO & \xNO & \xNO & \xPA & \xNO & \xNO  & \xNO    \\
		Edge-IIoTset~\cite{Ferrag2022}      & real, VM        & \xSI & \xSI & \xNO & \xNA & \xNO & \xSI & \xSI & \xNO & \xNO & \xNO & \xNO & \xNO & \xSI & \xSI & \xNO  & \xNO    \\
		MiniCPS~\cite{Antonioli2015}        & mininet         & \xNA & \xSI & \xNO & \xNA & \xSI & \xNO & \xNO & \xNO & \xSI & \xSI & \xSI & \xSI & \xSI & \xSI & \xSI  & \xNO    \\
		CPS\,Twinning~\cite{Eckhart2018}    & mininet         & \xNA & \xSI & \xNO & \xNA & \xSI & \xNO & \xNO & \xNO & \xSI & \xSI & \xSI & \xSI & \xSI & \xSI & \xSI  & \xNO    \\
		\textit{Gotham (Ours)}              & container, VM   & \xSI & \xSI & \xSI & \xSI & \xSI & \xSI & \xSI & \xSI & \xSI & \xSI & \xSI & \xSI & \xSI & \xSI & \xSI  & \xNO    \\
		ns-3, CupCarbon, $\ldots$ \cite{Chernyshev2018} & discrete event sim  & \xNO  & \xNO  & \xNO  & \xSI  & \xNO  & \xNO  & \xNO  & \xNO  & \xSI  & \xSI  & \xNA  & \xSI  & \xSI  & \xNO  & \xNA  & \xSI \\
		\bottomrule
	\end{tabularx}
	\vskip 1mm
	{\footnotesize \raggedright '\xSI': yes. \quad '\xPA': partially. \quad '\xNA': information not discussed or available or not applicable. \quad '\xNO': no. \hfill}
\end{table*}

To expand on the discussion presented in Section~\ref{sec:relatedwork}, Table~\ref{tab:compare_sota_feats} compares the cited testbeds/datasets according to the testbed property taxonomy outlined in this section (\figurename~\ref{fig:testbed_properties}). When a proposal does not meet a particular feature, it does not imply a fault in the testbed; however, it is insufficient for our needs to create a reproducible and flexible testbed. For instance, many testbeds are only focused on specific protocols \cite{Hindy2021, Vaccari2020, Hussain2021} and, thus, lack heterogeneity. While others include a wide variety of attacks, they lack fidelity because real malware activities are not included~\cite{Koroniotis2019, Moustafa2021, Ferrag2022} or vice versa~\cite{Meidan2018, GuerraManzanares2020}. Regarding \textbf{M1}, while all can capture network data, most only do it at specific choke points (port mirroring in a switch or router) instead of an arbitrary node. More importantly, most testbeds are unavailable and cannot be reproduced even if the main parts are mostly virtualized or containerized (the datasets created with them are available). Two of them~\cite{Antonioli2015, Eckhart2018} are reproducible and available; however, they are focused on PLC and HMI emulation, which differs from our proposal.

The physical lower-layer IoT communication protocols (e.g., Bluetooth, Zigbee, LoRa) are currently outside the scope of the testbed. This is further discussed in sections~\ref{sec:testbedimplementation} and~\ref{sec:discussion}.

\section{Testbed architecture} \label{sec:testbedarch}

The developed IoT network security testbed is based on the GNS3 network emulator~\cite{Grossmann}. GNS3 allows the creation of complex topologies composed of VMs, containerized images and real devices. It is actively developed and widely used in the industry and as a teaching tool for academia; the familiarity of this platform can ease the adoption of the proposed testbed. GNS3 is free software under the GNU GPLv3 license.

\begin{figure}[t]
	\centering
	\includegraphics[width=1.0\linewidth]{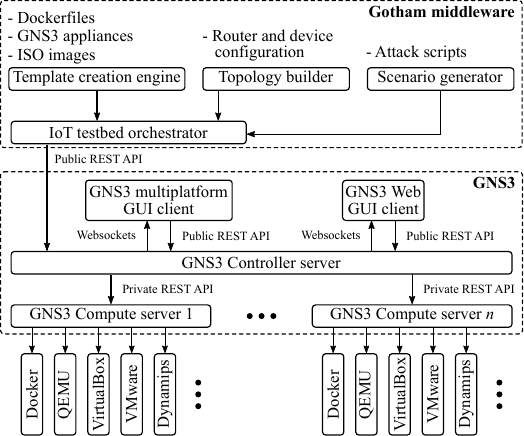}
	\caption{Gotham testbed architecture.}
	\label{fig:architecture}
\end{figure}

\figurename~\ref{fig:architecture} illustrates the proposed architecture, which includes the GNS3 components and the middleware built on top of it to implement the Gotham testbed. The central GNS3 component is the Controller server, which is responsible for managing all the projects, and it serves as an interface between the Clients and the Compute servers. The Clients allow the user to build the emulated network topology and interact with it by sending API requests to the Controller. The Compute servers are the software components that manage the different emulation engines supported by GNS3, such as Docker~\cite{Docker} containers, VMs based on QEMU~\cite{Bellard2005} or other hypervisors and Dynamips to emulate Cisco hardware. GNS3 allows running nodes in multiple compute server instances to achieve higher scalability~\cite{GNS3API}.

\subsection{Gotham middleware components}

Gotham is implemented as a GNS3 client and a set of programs that communicate with the GNS3 Controller via the public REST API. In the following, we describe the four components of the Gotham middleware shown in \figurename~\ref{fig:architecture}.


\subsubsection{IoT testbed orchestrator}

A set of functions that wrap around the GNS3 REST API~\cite{GNS3API} to automate and simplify various tasks such as node creation, node configuration (network interfaces, environment variables, executing configuration scripts, etc.), link creation, starting and stopping packet capturing in links and more. The rest of the components rely on these functions to build the topology and run the scenarios. They are executed in the following order: template creation engine, topology builder and scenario generator.

\subsubsection{Template creation engine}

Gotham uses QEMU VMs to emulate routers and Docker containers to emulate all the IoT nodes, attackers, servers and switches. The template creation engine builds all the Dockerfiles, sets up the ISO images of the VMs and generates GNS3 appliance templates representing those nodes. A template is a device model used to instantiate a node in the topology; GNS3 can create many nodes from a single template. GNS3 allows emulating networking equipment from multiple vendors; however, those images are usually proprietary and under licensing restrictions. We only include nodes based on free and open source software to ease reproducibility. Docker-based node templates include settings such as Docker image name, additional environment variables, start command and Docker volumes. The QEMU-based templates include settings like disk image files, RAM and CPU limits and other QEMU command-line parameters. After the template creation, the topology builder is executed.

\subsubsection{Topology builder}

The topology builder module describes the full topology of the scenario being emulated. After execution, it automatically instantiates all the nodes (based on the previously created templates), configures them and creates the necessary links to define the topology. The Docker-based images are configured by editing the \texttt{/etc/network/interfaces} file and setting the appropriate environment variables for each of them. In Gotham, the QEMU routers are configured from scratch by first installing the router operating system into the disk image and then configuring all interfaces and routing protocols for each VM. Once the topology has been defined, the scenario generator is executed to start the experiment.

\subsubsection{Scenario generator}

The scenario generator module starts all the nodes in a specific order and sets runtime options such as limiting the amount of memory and CPU quota a Docker container can use, setting bandwidth limits to network interfaces or starting and stopping packet capturing. Then, the scenario generator can schedule the launch of some attacks, or any other type of behavior, by running arbitrary scripts on the testbed nodes.

Regarding link and hardware resource emulation, currently, GNS3 has many features but also some limitations. Gotham addresses them in the following ways:

\paragraph{Link emulation} GNS3 allows modifying network link behavior by applying filters to packets in both directions, including packet dropping by frequency, packet loss percentage, delays, packet corruption percentage and filtering packets that match a Berkeley Packet Filter expression. However, applying bandwidth limits is not currently supported. To circumvent this limitation, we rely on \texttt{tc} (Linux Traffic Control)~\cite{Tcnetem} to provide a more realistic link emulation. In addition, GNS3 includes link status detection for QEMU-based nodes. When a link in the topology is suspended or removed, GNS3 will inform the node that the link status has changed, allowing a better router and routing protocol behavior emulation.

\paragraph{Hardware resource emulation} GNS3 only supports memory and CPU limits for nodes running in a hypervisor such as QEMU. Docker containers are not limited and can use all the available resources. However, to overcome this limitation, Gotham integrates the Docker API to apply memory and CPU constraints for resource emulation in containers.

\section{IoT scenario use case} \label{sec:testbedimplementation}

To illustrate the capabilities of the testbed, we have designed, implemented and validated the Gotham city scenario. An IoT use case scenario that contains multiple network segments, including building monitoring devices, domotics for a small neighborhood, industrial companies and malicious actors.

First, we outline the general network diagram of the scenario in terms of three layers: edge, network and cloud. Then, we detail the technical implementation regarding all the different emulated devices that run at each layer. Next, we describe the three threat models included in the scenario and the various attacks each can perform. Finally, we show the entire network topology of the scenario.

\subsection{Scenario diagram}

\begin{figure*}
	\centering
	\includegraphics[width=1.0\linewidth]{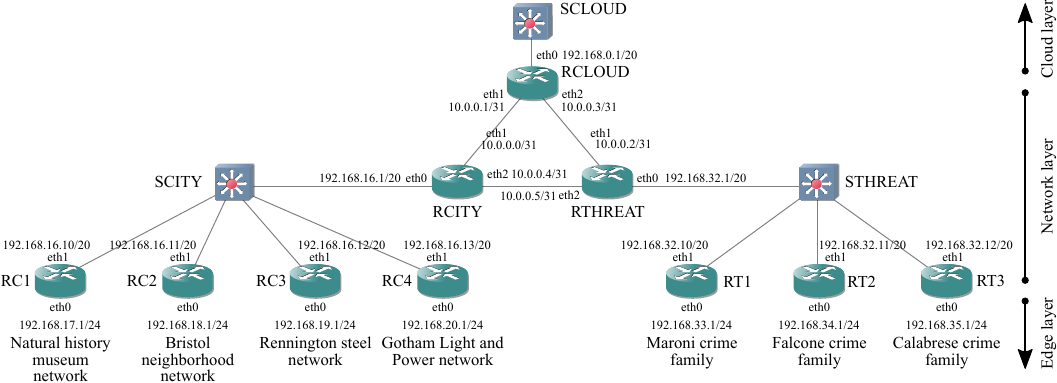}
	\caption{Network diagram for the emulated scenario.}
	\label{fig:scenario_network}
\end{figure*}

The network diagram for the scenario, including IP addresses and subnet masks for all interfaces, is shown in \figurename~\ref{fig:scenario_network}. For the purposes of the scenario, the 192.168.0.0/16 range is considered a publicly addressable range. The diagram only shows a partial view; the entire topology, including all the devices, servers and attackers, will be shown later in this section. The emulated scenario is divided into three main layers: edge, network and cloud layers.

\subsubsection{Edge layer}
The edge layer is composed of the emulated IoT/IIoT devices and attacker nodes. As shown in \figurename~\ref{fig:scenario_network}, the edge layer devices are located across two main zones: the city and the threat zones. The city zone devices include all the IoT and IIoT devices under the routers labeled RC1 to RC4. They communicate with the corresponding services at the cloud layer using different protocols and communication patterns. All the edge layer devices are addressable from any other node. These devices represent the publicly accessible devices located, for example, at the DMZ or outside the firewall of different industrial or residential networks of Gotham. The city zone is further divided into four segments. Each segment represents a different establishment generating traffic patterns based on realistic use cases:

\begin{itemize}
	\item \textit{Natural history museum}: A big building with many monitoring sensors and surveillance IP camera streams sending data to the cloud.

	\item \textit{Bristol neighborhood}: A group of houses that transmit data related to domotic systems, air quality measurements and IP camera streams.

	\item \textit{Rennington steel}: An industrial network sending telemetry data to the cloud for predictive maintenance purposes such as motor or tool failure event monitoring.

	\item \textit{Gotham Light and Power}: Another industrial network with IIoT nodes sending condition monitoring data from power generation plants and hydraulic test rigs.
\end{itemize}

The threat zone devices are located under the RT1, RT2 and RT3 router segments, each corresponding to a different threat actor, namely, the \textit{Maroni}, \textit{Falcone} and \textit{Calabrese} crime families:

\begin{itemize}
	\item \textit{Maroni crime family:} Portrays a threat model where external attackers scan and compromise IoT devices to turn them into bots. Includes a C\&C server and the supporting infrastructure for botnet propagation and launching attacks.

	\item \textit{Falcone crime family:} Depicts a threat model where legitimate IoT devices (in the city zone) have been previously compromised and maintain a connection with a C\&C server for remote control.

	\item \textit{Calabrese crime family:} Represents threats that externally scan IoT devices and launch attacks specifically targeting weaknesses in MQTT and CoAP protocols.
\end{itemize}

The details about the malicious activities and attacks are going to be described later in this section.

\subsubsection{Network layer}
The network layer devices are the switches and routers shown in \figurename~\ref{fig:scenario_network} that provide connectivity between the edge and cloud layers. Routers RCLOUD, RTHREAT and RCITY are the backbone routers of the testbed. They are configured with the OSPF routing protocol to update their routing tables dynamically. The edge layer routers from the city and threat zones are configured with static routing tables.

\subsubsection{Cloud layer}

The cloud layer includes the infrastructure that provides services to the edge layer devices. It includes additional services such as DNS and NTP. The cloud layer devices are connected to the RCLOUD router network.

\subsection{Emulated devices}

Here we provide the technical details referring to the implementation of the IoT/IIoT devices, attackers, network equipment and cloud infrastructure included in the scenario. Each node's implementation source code and artifacts are publicly available at~\cite{MiRepositorio}. The description is structured again in terms of edge, network and cloud layer devices.

\subsubsection{Edge layer devices}

Edge layer devices are responsible for generating the majority of the testbed's workload, including both legitimate and malicious network traffic. All the edge layer nodes are implemented as Docker containers and are fully reproducible thanks to Dockerfiles and the included dependencies, such as the programs implementing the node's behavior and the configuration files. The developed edge layer device templates, including IoT and attack nodes, can be instantiated multiple times in the testbed, and each instance can be configured differently to emulate distinct behavior patterns. The three main protocols used for IoT communication are MQTT, CoAP and RTSP.

The included IoT nodes represent devices located in urban or residential zones as well as IIoT devices for industrial equipment. The devices emulate IoT hubs or gateways, i.e., devices that connect to and gather data from various sensors, actuators or lower-level IoT devices and then communicate the collected data to the cloud or accept connections from external devices to query data or control the device. These types of IoT hubs and gateways that provide network layer connectivity are an integral part of IoT systems~\cite{Minerva2015}. The low-level connection between the emulated IoT hubs and the sensors (e.g., Bluetooth, Zigbee, etc.) is currently outside the scope of the testbed and not included; however, the data collection process is simulated in each IoT device by reading data obtained from multiple publicly available datasets related to specific IoT use cases. The data is used to generate a realistic-looking payload in terms of data volume and variety. The generated network traffic depends on how the emulated IoT devices transmit their payload, including the transmission protocol, periodicity, network conditions and interactions with other emulated devices defined in the scenario.

Regarding the data transmission behavior, we differentiate two modes: Open-close and Always-open. In Open-close, each time the device needs to send telemetry data, it opens a new connection to the cloud, sends the data and then closes the connection. In Always-open mode, the device opens a single connection with the cloud at the beginning and keeps it alive by periodically sending data and keep alive messages. Regarding the periodicity profiles, we also differentiate two modes: Continuous and Intermittent. In Continuous mode, the device is always actively sending telemetry data periodically. In Intermittent mode, the device has both active and inactivity time ranges. Active ranges work like the Continuous mode, but during inactivity ranges, no telemetry is transmitted, only background traffic.

\paragraph{MQTT-based edge devices}

Each MQTT-based device behavior is implemented in a Python program that uses the Eclipse Paho~\cite{Eclipsepaho} MQTT Python client library. The program's main thread periodically reads each line of the dataset, processes the data, builds the telemetry payload formatting it as a JSON, XML or Base64 encoded string and sends it to a broker using one or more topics in plain text or using TLS encryption. Furthermore, to increase the devices' heterogeneity and fidelity, the program includes other threads executing in parallel to create background traffic such as DNS, NTP requests and sending ICMP messages. The following lists the six MQTT-based edge layer device templates:

\textbf{Air quality}: Emulates an air quality chemical multisensor device based on the \cite{Airquality} dataset. It includes 15 sensor readings such as temperature, humidity and gas concentration sensors. The sensor data is transmitted as an XML payload of $\approx~1190$~bytes/record using a single MQTT topic in Continuous and Open-close mode.

\textbf{Building monitor}: Emulates a building condition monitoring based on the \cite{Buildingmonitor} dataset. It includes 27 humidity, temperature and energy consumption sensors located in different rooms. The sensor data is transmitted as a JSON payload of $\approx~100$~bytes/record using 11 MQTT topics in Continuous and Open-close mode.

\textbf{Cooler motor}: Emulates a device to monitor the vibration patterns of a fan based on the \cite{Coolermotor} dataset. It includes 5 sensors, such as acceleration and rotational speed sensors. The sensor data is transmitted as a Base64 encoded binary payload of $\approx~56$~bytes/record using a single MQTT topic in Intermittent and Always-open mode.

\textbf{Domotic monitor}: Emulates a monitoring system mounted in a domotic house based on the \cite{Domoticmonitor} dataset. It includes 24 sensors, such as wind, precipitation, CO2 concentration, and lighting. The sensor data is transmitted as an XML payload of $\approx~1743$~bytes/record using a single MQTT topic in Continuous and Open-close mode.

\textbf{Hydraulic system}: Emulates a device measuring process values from a hydraulic test rig based on the \cite{Hydraulicsystem} dataset. It includes 17 sensors measuring quantities such as pressure, power, flow, temperature and vibration. The sensor data is transmitted as a JSON payload with Base64 encoded values of $\approx~7678$~bytes/record using a single MQTT topic in Continuous and Always-open mode.

\textbf{Predictive maintenance}: Emulates a predictive maintenance system based on the \cite{Predictivemaintenance} dataset. It includes 14 sensors such as temperature, speed and torque for different product variants. The sensor data is transmitted as a JSON payload of $\approx~632$~bytes/record using 3 MQTT topics in Continuous and Open-close mode.

\paragraph{CoAP-based edge devices}
Each CoAP-based device implements a CoAP server using the libcoap C library~\cite{Libcoap}. The device creates a CoAP resource for each variable in the dataset, and it is served to clients at the cloud layer in plain text or using DTLS encryption. The clients periodically request data from the edge CoAP devices and perform other actions like resource discovery or sending ICMP messages. The following lists the two CoAP-based edge layer device templates:

\textbf{City power}: Emulates a city power consumption meter based on the \cite{Citypower} dataset. It includes 9 sensors, such as power consumption in three city zones and weather information. Each sensor data size is $\approx~10$~bytes and serves 9 CoAP resources.

\textbf{Combined cycle}: Emulates the monitoring of a combined cycle power plant based on the \cite{Combinedcycle} dataset. It includes 5 sensors: temperature, pressure, humidity, exhaust vacuum and energy output. Each sensor data size is $\approx~10$~bytes and serves 5 CoAP resources.

\paragraph{RTSP-based edge devices}
The devices based on RTSP are the IP cameras and IP camera stream consumers. The IP cameras send a looped video file through the network using FFmpeg~\cite{FFmpeg} to an RTSP server at the cloud layer. The stream consumers read the video feed from the server also using FFmpeg. They use a variety of protocols, including RTP, RTCP, RTSP, ICMP and DNS. The following lists the three camera edge layer device templates:

\textbf{IP camera (x2)}: Includes 2 templates using different video files and settings: Video 1, adapted from \cite{VideoLondon}, sends 1280x720 resolution, 15 fps, color, no audio, libx265 codec, 40s looped stream. Video 2, adapted from \cite{VideoLebanon}, sends 1280x720 resolution, 25 fps, grayscale, no audio, libx264 codec, 16s looped stream. On average, the Video 1 stream generates $\approx~1.2$~Mbit/s traffic, and Video 2 generates $\approx~1.8$~Mbit/s traffic. They write to the stream server in Continuous mode.

\textbf{IP camera consumer}: Emulates a system reading from a video stream generated by an IP camera. Each stream generates $\approx~1.8$~Mbit/s traffic. It reads from the stream server in Intermittent mode.

For comparison with real IoT IP cameras, in~\cite{Mirsky2018}, the authors show a table with specifications and statistics of some real IoT IP cameras used in their experiments. The specifications are close to our emulated camera nodes. They communicate with RTSP/RTP protocol, use the same image resolution, H.264 codec and 15 fps framerate. They generate around~$1.4$~--~$1.8$~Mbit/s traffic on average, similar to our emulated cameras.

\paragraph{Attacker or malicious edge devices}
Regarding the attacker or malicious nodes, the scenario includes the following ten templates:

\textbf{Mirai bot}: This device includes the Mirai bot binary adapted and compiled from~\cite{Miraisrc}. After execution, the bot can perform several steps: network scanning, brute force authentication, reporting gathered credentials to the scan listener server and performing multiple DoS attacks.

\textbf{Mirai C\&C}: This device is composed of a MySQL database, and the Mirai Command \& Control server adapted and compiled from~\cite{Miraisrc}. With the appropriate username and password, clients can connect to the C\&C to schedule DDoS attacks. The database holds credentials, client information and attack records.

\textbf{Mirai scan listener}: This is the Mirai scan listener binary compiled from~\cite{Miraisrc}. The server listens to the bot scanning results when a successful Telnet username and password have been found.

\textbf{Mirai loader}: This device includes the main Mirai loader binary adapted and compiled from~\cite{Miraisrc} and Mirai downloader binaries for 9 architectures. The loader program takes the scanning reports from the scan listener server and logs into each device to download and execute the Mirai bot.

\textbf{Mirai download server}: It is an HTTP server hosting the Mirai bot binary.

\textbf{Merlin C\&C}: This device includes the Merlin cross-platform post-exploitation Command \& Control server~\cite{Tuyl}. All the devices compromised with the Merlin agent report to the Merlin C\&C. Users can connect to this node to control each bot.

\textbf{Scanner}: The scanner node contains the Nmap~\cite{Nmap} and Masscan~\cite{Masscan} network scanners. Additionally, Nmap can probe MQTT brokers.

\textbf{MQTT attacks}: The node includes the SlowTT~\cite{Slowttattack} and MQTTSA~\cite{Mqttsa} tools to perform attacks against MQTT.

\textbf{CoAP attacks}: The node includes the AMP-Research~\cite{Ampresearch} tool to perform amplification attacks against CoAP devices.

\textbf{Metasploit}: It includes the Metasploit Framework~\cite{Metasploit} for executing exploit codes against remote targets.

\paragraph{Edge device configuration}
Each device instance created from the described templates can be configured by setting environment variables for the Docker containers. These variables are set in the topology creation program. The configuration options include the MQTT broker address or domain name, CoAP server address, MQTT topic, MQTT username and password authentication, MQTT QoS values, enabling TLS for MQTT or DTLS for CoAP, sleep times (mean value and random standard deviation) for each thread or setting the active and inactive time periods to enable and disable the telemetry thread temporarily. Additionally, all the IoT devices include the BusyBox~\cite{Busybox} binary. To make some devices vulnerable to Mirai, the scenario generator program configures some nodes with a username and password combination found in Mirai's brute forcing table, running the BusyBox Telnet server and setting the login shell to the BusyBox shell.

\subsubsection{Network layer devices}

\begin{table}[t]
	\small
	\centering
	\caption{Network Layer Device Templates.}
	\label{tab:netlyr_devices}
	
	\begin{tabularx}{1.0\linewidth}{p{0.27\linewidth}|X}
		\toprule
		\textbf{Device} & \textbf{Description} \\
		\midrule
		Switches & Docker image with Open vSwitch~\cite{Openvswitch} version 2.12.3. Each switch has 16 network interfaces. \\
		Backbone routers: {\footnotesize RCLOUD, RTHREAT, RCITY}  & QEMU VM with VyOS~\cite{Vyos} version 1.3.0-rc6. OSPF protocol is used to configure the routing tables dynamically. \\
		Edge routers: {\footnotesize RC1--5, RT1--3} & QEMU VM with VyOS~\cite{Vyos} version 1.3.0-rc6. The routers are configured using static routing tables and Proxy ARP is enabled. \\
		\bottomrule
	\end{tabularx}
\end{table}

The information about the switches and routers used in the testbed is summarized in Table~\ref{tab:netlyr_devices}. The configuration of all the routers is performed automatically by the topology creation program. After completing the image installation, each router is configured by executing a Bash script with all the necessary VyOS CLI commands. All routers are provided with 512MB RAM, 1 virtual CPU, KVM acceleration and the virtio para-virtualized network adapter to increase the performance of the network adapters.

\subsubsection{Cloud layer devices}

\begin{table}[t]
	\small
	\centering
	\caption{Cloud Layer Device Templates.}
	\label{tab:cloudlyr_devices}
	
	\begin{tabularx}{1.0\linewidth}{p{0.21\linewidth}|X}
		\toprule
		\textbf{Device template} & \textbf{Description} \\
		\midrule
		MQTT broker plain text & Docker with Eclipse Mosquitto~\cite{Light2017} 1.6 in its default configuration. \\
		MQTT broker plain text with authentication & Docker with Eclipse Mosquitto~\cite{Light2017} 1.6 configured with 2 username/password combinations to only accept authenticated clients. \\
		MQTT broker TLS encryption & Docker with Eclipse Mosquitto~\cite{Light2017} 2.0 configured with X.509 certificates to enable encryption. \\
		Combined cycle CoAP client & Ubuntu Docker image with a libcoap~\cite{Libcoap} client requesting services provided by the Combined cycle IoT servers from the edge layer. Can communicate in plain text or encrypted using DTLS with pre-shared keys. \\
		City power CoAP client & Ubuntu Docker image with a libcoap~\cite{Libcoap} client requesting services provided by the City power IoT servers from the edge layer. Can communicate in plain text or encrypted using DTLS with pre-shared keys. \\
		IP camera stream server & Alpine docker image. RTSP-simple-server~\cite{Rtspsimpleserver} is installed and configured. \\
		DNS & Ubuntu Docker image. Dnsmasq~\cite{Dnsmasq} is installed and configured to provide DNS services. \\
		NTP & Alpine Docker image. Chrony~\cite{Chrony} is installed and configured to provide NTP services. \\
		\bottomrule
	\end{tabularx}
\end{table}

The cloud layer comprises multiple MQTT brokers running differently configured instances of the Eclipse Mosquitto~\cite{Light2017} broker, CoAP clients based on the libcoap library and IP camera streaming servers running the RTSP-simple-server~\cite{Rtspsimpleserver} that allows clients to publish and read audio and video streams. Additionally, it includes DNS and NTP services used by most edge layer devices. The topology creation program automatically configures the testbed-specific configuration parameters, such as device addresses or DNS hosts and names. A description of the cloud layer device templates is shown in Table~\ref{tab:cloudlyr_devices}.

\subsection{Threat model and attacks} \label{sec:threatmodelattacks}

Here we detail the three threat actors included in the scenario, portrayed by three crime families in Gotham: \textit{Maroni}, \textit{Falcone} and \textit{Calabrese}. Each threat actor models different malicious activities, which in conjunction, represent a comprehensive threat model to the IoT.

First, \textit{Maroni} represents external attackers that perform automated actions to scan, exploit and control IoT devices. Then, \textit{Falcone} represents previously compromised IoT devices by an unknown method (e.g., by insiders, manufacturers or supply chain attacks) that connect to an external network controlled by the attacker. Finally, \textit{Calabrese} represents attacks specifically targeting some IoT protocol weaknesses.

While we include a diverse set of activities in this particular scenario, the testbed scenario can be extended to include even more attacks depending on the interest of researchers or to test new attacks as they are discovered.

\subsubsection{Maroni Crime Family}

This threat actor represents an attacker-controlled network that remotely scans, attacks and compromises other devices to incorporate them into a botnet. The devices in this threat actor include the four Mirai nodes in the RT1 network and the single bot located at the cloud layer. All the nodes are based on binaries compiled from the published Mirai source code~\cite{Miraisrc}.

To adapt the malware to the closed testbed environment, several modifications had to be made to the Mirai source code. All the modifications and details are available in the testbed repository~\cite{MiRepositorio}. Briefly, the changes include: (i) replacing the hardcoded DNS address (8.8.8.8) with the address of the DNS server in the lab, (ii) replacing the hardcoded loader address with the one used in the lab, (iii) removing C preprocessor directives to enable port scanning and launching attacks when compiling in debug mode (to be able to see Mirai log messages), (iv) patching the function to generate random IP addresses to include the ranges used in the testbed and other minor changes such as (v) using Unix sockets instead of TCP to open the database locally and (vi) fixing some errors in the database creation scripts. The following is a list of the included malicious behavior:

\begin{itemize}
	\item \textbf{Periodic C\&C communication:} Mirai bots perform periodic communication with both the C\&C server and the loader server.
	
	\item \textbf{Network scanning:} Each Mirai bot scans the network in a pseudorandom order sending TCP SYN packets to the 23 and 2323 ports.
	
	\item \textbf{Brute forcing:} When a potential victim is found during the network scanning phase, the bot tries to brute force the victim's Telnet credentials using a hardcoded list of username and password combinations.
	
	\item \textbf{Reporting:} After a successful brute forcing, the Mirai bot sends the victim's IP address, port, username and password combination to the Mirai scan listener server.
	
	\item \textbf{Ingress tool transfer:} For each vulnerable device listed in the listener server, the loader program logs in and downloads the malware into each device.
	
	\item \textbf{Remote command execution:} The Mirai C\&C server can instruct the bots to launch various attacks against the victims.
	
	\item \textbf{Denial of service attacks:} Mirai includes 10 DoS attack types, including network and application layer attacks: generic UDP flood, UDP flood optimized for higher speeds, flood against game servers running the Valve Source engine, DNS flood, TCP SYN, TCP ACK attacks, TCP stomp flood, GRE IP flood, GRE Ethernet flood and HTTP flood.
\end{itemize}

For a more detailed description of Mirai's behavior, operation, lifecycle and attacks, please refer to~\cite{Kambourakis2017, Antonakakis2017}.

\subsubsection{Falcone Crime Family}

This threat actor represents a set of devices that have been previously compromised but are still running by legitimate users inside the city zone network. The infection vector is not relevant in this case; this could represent, for instance, supply chain attacks, malware installed by the manufacturer or insider attacks. The attacker-controlled node is a single Docker container running the Merlin~\cite{Tuyl} multi-platform post-exploitation Command and Control (C\&C) server, and the compromised devices run the Merlin agent. The Merlin server supports multiple protocols for C\&C (HTTP/1.1 clear-text, HTTP/1.1 over TLS, HTTP/2, HTTP/2 clear-text, HTTP/2 over QUIC) and can remotely execute arbitrary code on the bots under its control. To generate attacks against the victims, the node also includes the hping3~\cite{Sanfilippo} TCP/IP packet assembler and analyzer.

To increase the variety within the same attack categories, the currently implemented attacks for this threat actor are DoS-based attacks similar to Mirai's attacking behavior as described in its source code~\cite{Miraisrc} but implemented using hping3. However, since the Merlin C\&C allows the execution of arbitrary commands in the controlled machines, it can be used as a generic tool to perform various attacks. The following is a list of the currently performed malicious behavior and attacks:

\begin{itemize}
	\item \textbf{Periodic C\&C communication:} The Merlin C\&C server is initialized and starts listening for incoming connections. All the compromised nodes execute the Merlin agent and connect to the server. The clients periodically communicate with the server to keep alive the C\&C channel.
	
	\item \textbf{Ingress tool transfer:} The C\&C server transfers the hping3 binary into each of the compromised devices. The tool is used to perform subsequent DoS attacks against other targets in the network.
	
	\item \textbf{Remote code execution:} The server remotely executes commands into the compromised machines to prepare the environment for the previously uploaded hping3 binary.
	
	\item \textbf{Denial of service attacks:} The Merlin C\&C server commands the compromised devices to send various flooding attacks against a selected target. The attacks include sending ICMP echo requests, UDP generic flood to different random ports, TCP SYN and TCP ACK attacks.
\end{itemize}

\subsubsection{Calabrese Crime Family}

This threat actor first performs network-wide scanning activities to identify all the IoT devices in the testbed as a precursor to launching targeted attacks against the devices that use the MQTT and CoAP protocols. The scanning is performed by the Scanner node, which includes two different tools: Nmap~\cite{Nmap} and Masscan~\cite{Masscan}. Nmap can additionally be used to establish a connection to an MQTT broker to listen and read all the messages being published by the clients. Attacks against the MQTT broker are implemented using the MQTT attacks node, which includes the MQTTSA~\cite{Mqttsa} and SlowTT-Attack~\cite{Slowttattack} tools, and the Metasploit node, which includes the Metasploit Framework~\cite{Metasploit}. The attack against CoAP nodes is implemented using code provided by AMP-Research~\cite{Ampresearch}. The malicious behavior performed by this threat actor includes:

\begin{itemize}
	\item \textbf{Network scans:} Performs network-wide scans to identify and collect information about all the available hosts in the testbed and the different services they are running. It uses Nmap and Masscan.
	
	\item \textbf{MQTT sniffing attack:} Intercepts MQTT connect packets and searches for credentials to connect to the MQTT broker. It uses the MQTTSA tool.
	
	\item \textbf{MQTT brute force:} Uses multiple wordlists containing common usernames and passwords to brute force login credentials to the broker. It uses MQTTSA and Metasploit for the attack, with wordlists also provided by Metasploit.
	
	\item \textbf{MQTT information disclosure:} For unauthenticated brokers, or after discovering the credentials by sniffing or brute forcing, it reads all the messages being published to an MQTT broker by subscribing to all data topics (\#), control topics (\$SYS) or only some specific topics. It uses Nmap and MQTTSA tools.
	
	\item \textbf{MQTT malformed data:} Sends malformed packets to the broker in order to trigger exceptions caused by errors in the server's input validation methods. It uses the MQTTSA tool.
	
	\item \textbf{MQTT denial of service:} First, it includes a \textit{slow} DoS attack by creating a high number of parallel connections to the broker and keeping them alive indefinitely. Then, it saturates the broker by publishing large payloads with many clients. It uses MQTTSA and SlowTT-Attack tools.
	
	\item \textbf{CoAP amplification attack:} The attacker sends a small request to a CoAP server that generates a response payload larger than the request. The attacker can abuse this by spoofing the source address, causing the response to be directed to a victim. If this attack is rapidly repeated, it can cause a denial of service on the victim, and since the traffic is reflected using legitimate servers, it can be challenging to block by using simple blocklists. It uses the AMP-Research tool.
\end{itemize}

The increase in threats that specifically target exposed MQTT and CoAP devices has been raised in an industry report in which they scan and find numerous exposed and vulnerable production systems and show how those attacks operate~\cite{Maggi2018}.

\subsection{Full network topology}

\begin{figure}[t]
	\centering
	\includegraphics[width=1.0\linewidth]{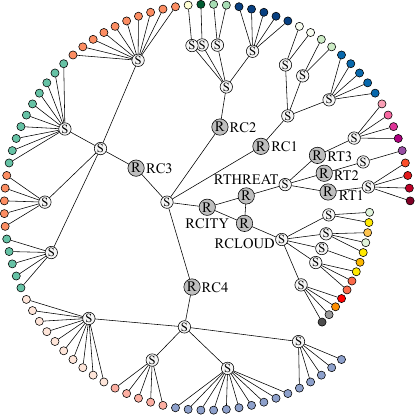}
	\caption{The full network topology of the scenario as an undirected graph. $\circledR$ and $\circledS$ represent routers and switches, respectively, and the colored circles represent different instances of edge and cloud layer devices.}
	\label{fig:topograph}
\end{figure}

The complete topology of the emulated scenario, including all instances of edge, network and cloud devices, is shown in \figurename~\ref{fig:topograph}. There are 100 edge and cloud devices, 30 switches and 10 routers.

The devices at the cloud layer include: 1 DNS, 1 NTP, 1 Mirai bot, 5 MQTT brokers (3 plain text, 1 with authentication and 1 with TLS), 2 stream servers, 1 City power CoAP client and 2 Combined cycle CoAP clients (plain text and with DTLS encryption).

Regarding the threat zone, the devices connected to RT1 include the Mirai-related nodes except for the Mirai bot (located at the cloud layer), for a total of 4 devices. RT2 includes 1 Merlin C\&C server, and RT3 includes 1 Scanner, 1 MQTT attacker, 1 CoAP attacker and 1 Metasploit node.

The city zone includes most of the devices. The natural history museum network (RC1) comprises 5 Building monitors communicating in plain text and without authentication, 2 IP cameras and 2 IP camera consumers. Bristol neighborhood (RC2) contains 5 Domotic monitors transmitting in plain text and without authentication, 2 IP cameras, 1 Air quality and 1 City power. The Rennington steel network (RC3) is composed of 15 Cooler motor nodes (10 communicating in plain text but with authentication and 5 using TLS), 15 Predictive maintenance nodes (10 in plain text with authentication and 5 encrypted with TLS). The Gotham Light and Power network (RC4) includes 15 Combined cycle nodes (10 communicating in plain text without authentication, 5 encrypted with DTLS), 15 Hydraulic system nodes (10 in plain text and without authentication and 5 with TLS).

\section{Evaluation} \label{sec:evaluation}

This section presents the discussion to validate the features from \figurename~\ref{fig:testbed_properties} based on the design of the testbed, the behavior of the included nodes and various experiments.

\subsection{Reproducibility}

As mentioned in sections~\ref{sec:testbedarch} and~\ref{sec:testbedimplementation}, the template creator builds node templates using Dockerfiles that describe all the dependencies, configuration variables and behavior in a reproducible way (\textbf{R1}). To provide reproducible attack scripts (\textbf{R2}), the scenario generator can launch attacks by running programs in arbitrary nodes. The network topology builder is a program that automatically reproduces the scenario topology (\textbf{R3}) after its execution. The use of formal languages to describe devices, behavior and topology allows an automated and reproducible setup to replicate the scenario.

\subsection{Communication link emulation}

IoT devices may be connected to the network via links of varying quality that should be emulated in the testbed. Here we are evaluating the link emulation fidelity for TCP and UDP traffic in terms of bandwidth shaping. We first select two edge layer nodes connected through a network switch from the testbed. The first experiment consists of limiting the network rate of the first node and measuring the maximum bitrate using the iPerf3~\cite{Iperf3} TCP test. One of the nodes runs iPerf3 in server mode, while the other runs it in client mode. The client limits the link rate using \texttt{tc} to 1, 10, 25, 50, 75 and 100 Mbit/s~\footnote{For example: \texttt{tc qdisc add dev eth0 root netem rate 100mbit}}. iPerf3 runs for 10 seconds for each rate limit, and each measure is repeated 20 times. The results are shown in Table~\ref{tab:iperf3tcp} with the measured mean bitrate, error and the coefficient of variation.

\begin{table}
	\small
	\centering
	\caption{Link Emulation Fidelity Using iPerf3 in TCP Mode.}
	\label{tab:iperf3tcp}
	
	\begin{tabularx}{1.0\linewidth}{X|X|X|X}
		\toprule
		\textbf{Rate limit} & \textbf{Measured} & \textbf{Error (\%)} & \textbf{CV (\%)} \\
		\midrule
		100 Mbit/s & 93.739 & 6.261 & 0.695 \\
		75 Mbit/s  & 71.502 & 4.664 & 0.501 \\
		50 Mbit/s  & 47.757 & 4.486 & 0.755 \\
		25 Mbit/s  & 23.867 & 4.532 & 0.379 \\
		10 Mbit/s  & 9.551 & 4.490 & 0.296 \\
		1 Mbit/s   & 0.957 & 4.300 & 0.044 \\
		\bottomrule
	\end{tabularx}
\end{table}

\begin{figure}
	\centering
	\includegraphics[trim=3mm 3mm 3mm 0mm,clip]{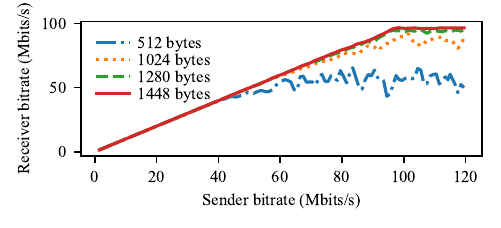}
	\caption{Link emulation fidelity using iperf3 sending udp data to a 100Mbits/s limited node.}
	\label{fig:iperf3udp}
\end{figure}

In the second experiment, iPerf3 UDP traffic performance is tested by sending UDP data at fixed bandwidths from the first node and measuring it at the receiver end. The link rate from the sender node to the switch is limited to 100~Mbit/s. The methodology is similar to the network performance test done in~\cite{Siaterlis2013}. The generated traffic ranges from 1~Mbit/s up to 120~Mbit/s, and it is repeated for UDP packet sizes from 512~bytes to 1448~bytes. The results are shown in \figurename~\ref{fig:iperf3udp}. For large packet sizes, the curve resembles the ideal behavior: a line with a slope of one up to the network link maximum rate limit and a horizontal line for faster bitrates. There are significant packet losses for smaller packet sizes before reaching the 100~Mbit/s link limit; the saturation point for each packet size depends on the hardware running the GNS3 emulator. However, the traffic rates generated by the emulated edge layer nodes all fall in the linear region of the curve.

Using the built-in GNS3 link emulation tools and installing \texttt{tc} into all the nodes to increase the link emulation features in the proposed testbed, the (\textbf{F4}) feature is satisfied.

\subsection{Hardware resource emulation}

\begin{figure}
	\centering
	\includegraphics[trim=3mm 3mm 3mm 3mm,clip]{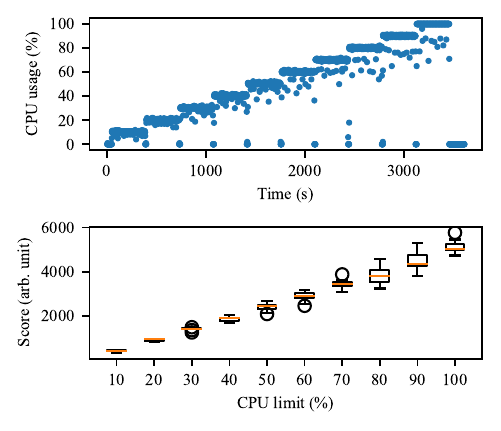}
	\caption{Hardware resource emulation fidelity. Top: actual CPU usage for varying CPU constraints. Bottom: stress-ng CPU benchmark scores under different CPU constraints.}
	\label{fig:cpulimits}
\end{figure}

Due to hardware heterogeneity, IoT devices can vary greatly in terms of computational capabilities. To evaluate the fidelity of hardware resource emulation of Docker-based nodes, we run the stress-ng~\cite{Stressng} tool inside a container while imposing CPU constraints using the Docker API. The container is limited to a single core, and the available CPU resources shared with the container are limited from 10\% to 100\% in 10\% increments. For each CPU resource limit, the stress-ng runs multiple CPU stressing methods for 30~seconds and repeated 11~times. \figurename~\ref{fig:cpulimits} (top) shows the total CPU usage of the core where the container is pinned for the entire duration of the experiment. The staircase pattern shows that the container does not use more CPU resources than the imposed limit. The slight increment is due to other processes outside the container running in the same CPU core. \figurename~\ref{fig:cpulimits} (bottom) depicts the obtained stress-ng benchmarking scores for each CPU limit. The exact value of the score is hardware dependent; however, it clearly shows a linear relationship between the benchmarking scores and CPU limits. The testbed can emulate different hardware resources (\textbf{F1}) using the provided features by the Docker engine.

\subsection{Testbed scalability}

\begin{figure}
	\centering
	\includegraphics[trim=3mm 3mm 3mm 3mm,clip]{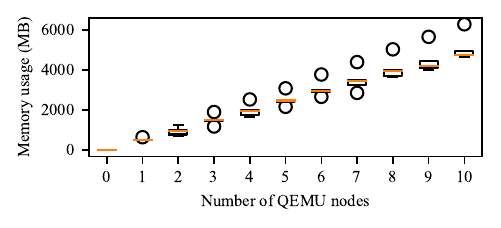}
	\caption{Memory scalability for QEMU nodes.}
	\label{fig:qemumem}
\end{figure}

\begin{figure}
	\centering
	\includegraphics[trim=3mm 3mm 3mm 3mm,clip]{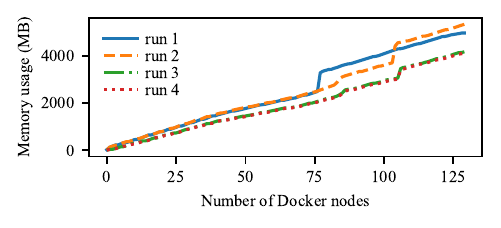}
	\caption{Memory scalability for Docker nodes.}
	\label{fig:dockermem}
\end{figure}

The scalability is measured in terms of the memory consumption required to instantiate all the scenario nodes from \figurename~\ref{fig:topograph}. The additional memory usage as a function of the number of running QEMU node instances (VyOS routers) is shown in \figurename~\ref{fig:qemumem}. To measure the memory usage for the Docker-based nodes, a new node is started every 5 seconds, beginning from the switches, followed by the nodes acting as servers, and finally, the rest of the nodes (\figurename~\ref{fig:dockermem}). The jumps in memory are due to cached memory. Both figures show a linear trend for memory usage, which allows estimating memory requirements depending on the desired scale of the scenario. In total, for the scale presented in the scenario from \figurename~\ref{fig:topograph}, slightly more than 10Gb of memory is required, allowing the emulation of medium to large-scale deployments even on a single machine (\textbf{S1}).

The presented Gotham testbed scenario is currently implemented using a single GNS3 instance running on a single machine, which can be a limiting factor in emulating scenarios with thousands of nodes based on the scalability measures shown here. However, GNS3 is not restricted to a single instance. Multiple GNS3 servers can be connected to one another through a physical network that acts as one large network to achieve higher scalability to create even larger scenarios.

\subsection{Measurability}

GNS3 allows packet capturing on any link connected between nodes of any type (independent of the underlying emulation engine) (\textbf{M1}). The user can also save artifacts (log files, binaries, etc.) (\textbf{M2}) by connecting to any node, allowing the creation of datasets that mix network and host-level data sources. The following experiments about normal and attack scenarios show examples of measurability.

To generate datasets from the testbed, the user can define in the scenario generator module all the links where the network traffic will be captured. To extract device log data, the user can leverage the docker API to get arbitrary files or command output inside any container, which can also be automated at the scenario generator script. The data capturing can be started and stopped at any time. All raw network traces and logs generated by the testbed can then be processed depending on the use case.

\subsection{Normal IoT behavior scenario}

This scenario allows us first to check the ability to deploy and run multiple devices communicating with different protocols and, secondly, to capture and verify network traffic data measurements. We start all the nodes included in the scenario (\figurename~\ref{fig:topograph}) except for the attackers. As explained in Section~\ref{sec:testbedimplementation}, to ensure device behavior fidelity (\textbf{F2}), the nodes run real production libraries to send the telemetry and also create diverse background traffic, including ICMP, DNS and NTP requests. The number of emulated devices and networking equipment included in the scenario allows the deployment of a sufficiently complex network topology that meets the (\textbf{F5}) feature. Additionally, the inclusion of multiple services with different configurations in the topology complies with the (\textbf{H3}) feature.

To verify the generated protocol diversity, we capture network traffic data for one hour at the link between RCLOUD and SCLOUD, and use Wireshark's~\cite{Wireshark} dissectors to identify the list of protocols and packet volume, as shown in Table~\ref{tab:protocols}. The traffic related to the IP cameras and stream consumer devices generate the largest number of packets due to the high volume of data transmitted compared to the devices using lighter MQTT and CoAP protocols. The currently included devices generate a varied protocol diversity, which satisfies (\textbf{H1}) for the purposes of the scenario.

\begin{table}[t]
	\small
	\centering
	\caption{List of Identified Protocols After One Hour of Normal IoT Traffic Capture (No Attacks) at the Link Between RCLOUD and SCLOUD.}
	\label{tab:protocols}
	
	\begin{minipage}{0.48\linewidth}\centering
		\begin{tabularx}{1.0\linewidth}{l|r}
			\toprule
			\textbf{Service} & \textbf{Packets (\%)} \\
			\midrule
			rtp & 87.638 \\
			tcp &  5.384 \\
			mqtt &  2.581 \\
			tls &  1.728 \\
			dns &  1.533 \\
			ntp &  0.278 \\
			rtcp &  0.265 \\
			\bottomrule
		\end{tabularx}
	\end{minipage}
	\hfill
	\begin{minipage}{0.48\linewidth}\centering
		\begin{tabularx}{1.0\linewidth}{l|r}
			\toprule
			\textbf{Service} & \textbf{Packets (\%)} \\
			\midrule
			dtls &  0.180 \\
			icmp &  0.139 \\
			coap &  0.121 \\
			arp &  0.120 \\
			rtsp &  0.020 \\
			icmpv6 &  0.009 \\
			sdp &  0.002 \\
			\bottomrule
		\end{tabularx}
	\end{minipage}
\end{table}

\subsection{Attack behavior scenario}

\begin{figure}[t]
	\centering
	\includegraphics[trim=3mm 3mm 3mm 3mm,clip]{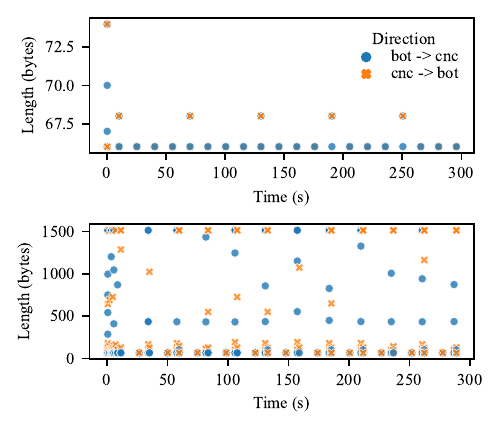}
	\caption{Periodic network packet communication between a bot and its C\&C server. Top: Mirai. Bottom: Merlin.}
	\label{fig:mirai-merlin-cnc}
\end{figure}

\begin{figure}[t]
	\centering
	\begin{minipage}{0.493\linewidth}\centering
		\includegraphics[width=1.0\linewidth,trim=0 27mm 0 0.5mm,clip]{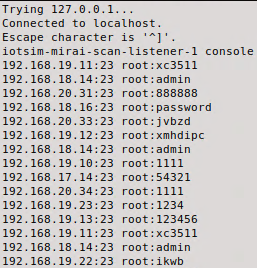}
	\end{minipage}
	\hfill
	\begin{minipage}{0.493\linewidth}\centering
		\includegraphics[width=1.0\linewidth,trim=0 0 0 27.5mm,clip]{images/saezdecamara-fig10.png}
	\end{minipage}
	\caption{Mirai scan listener reports.}
	\label{fig:miraiscanlistener}
\end{figure}

Here we validate the ability to execute attacks from the included threat models in the scenario, and measure the generated network traffic and logs. The attack scenario is prepared by making 24 edge layer nodes vulnerable to Mirai (setting appropriate usernames and passwords, starting BusyBox telnet server and changing the login shell to BusyBox sh; all automatically performed by the scenario generator module) and running the Mirai bot from the \textit{Maroni} threat actor. Another node is compromised by installing the Merlin bot agent and running the Merlin C\&C from the \textit{Falcone} threat actor. After executing the Mirai bot and the Merlin agent programs, the periodic communications with their respective C\&C can be observed in \figurename~\ref{fig:mirai-merlin-cnc} (top) for Mirai and \figurename~\ref{fig:mirai-merlin-cnc} (bottom) for Merlin. After the connection with the C\&C is established, the Mirai bot starts the scanning phase. \figurename~\ref{fig:miraiscanlistener} shows some successful brute forcing attempts reported to the Mirai scan listener node. Besides the network traffic, indications of the Mirai bot activity can be found, for instance, by inspecting the DNS logs inside the DNS server node. At this stage, the user can interact (manually or programmatically) with the corresponding C\&C servers to perform the attacks described in Section~\ref{sec:threatmodelattacks}.

Regarding the \textit{Calabrese} threat actor, Nmap and Masscan are used to perform both horizontal and vertical scans to any network in the testbed. Attacks can be launched against the identified MQTT brokers using the MQTT attacks, Metasploit and Scanner nodes. For instance, a successful brute forcing attack can be performed against the authenticated MQTT broker using the word lists provided by the Metasploit node. The CoAP attack node can leverage any CoAP server in the testbed to launch an amplification attack against a victim. The CoAP amplification attack sends a GET request to the \texttt{.well-known/core} resource with a spoofed source address to the server, which generates a response with a bigger payload directed to the victim. The request generates a 21 bytes CoAP payload and response of around 430 bytes (depending on the server), implying an amplification factor of approximately 20.

The included real botnet and the red-teaming tools can be launched against various targets in the testbed (\textbf{F3}). Also, by including overlapping tools that perform similar attacks using different implementations, the attack behavior diversity (\textbf{H2}) is achieved.


\section{Discussion} \label{sec:discussion}

In this work, we present the Gotham testbed, a security testbed that builds upon the GNS3 network emulator to provide a reproducible and flexible testbed that allows the creation of security scenarios to test attacks, defenses or extract datasets for ML model training. To generate real network traffic, we are leveraging QEMU-based VMs and Docker-based containerization technology to implement a scenario composed of emulated IoT/IIoT devices, servers and network equipment that run real production libraries, network switching software and routing operating systems as well as real malware samples. The implemented scenario comprises more than 30 different emulated device templates. The topology definition, creation and execution consist of several scripts that automatically instantiate and configure 140 nodes and set various runtime options, such as hardware limits and network link shaping.

The presented testbed has some potential limitations and considerations arising from the architectural design choices and the way in which scenarios are created. Currently, GNS3 does not directly support the emulation of wireless physical links and protocols, which can limit its use for low-power wireless sensor network security research. Similarly, while the simulation of IoT node mobility is not directly supported in GNS3, the Gotham testbed can simulate network quality that varies over time by periodically changing network link properties for certain nodes using the scenario generator script. To overcome these limitations, future work can explore the integration of the Gotham testbed with other lower-level network simulators with wireless simulation capabilities, such as ns-3.

Another consideration regarding the creation of different scenarios is the number of configuration steps a user needs to perform to integrate new IoT devices, servers, network equipment or malware nodes. The configuration includes steps such as configuring routers, modifying the source code of legitimate or malware applications, recompiling them, and creating Docker images. Most of the nodes included in the scenario described in this paper are independent and can be directly reused for different scenarios. However, other nodes can show a higher coupling between the node's behavior and the scenario. For example, the Mirai binary in the Mirai bot node includes some hardcoded values that are scenario specific. Nevertheless, due to the reproducibility property of the Gotham testbed, those nodes can be rebuilt to adapt them to different scenarios with minimal configuration changes. However, for this to be possible, a user creating new nodes must be careful and use good practices to maintain the reproducibility property and create flexible nodes that can run under different scenarios.

Regarding security considerations of the testbed itself, the user should be aware that, by default, GNS3 runs Docker containers in privileged mode. This detail could open the way for Docker-aware malware to escape the container. In such cases, or when unknown malware binaries are to be integrated into the scenario topology, the user should carefully consider the emulation engine to run the node to properly contain the malware, for example, using QEMU-based VMs instead of Docker containers and additionally hardening or isolating the machine(s) where the GNS3 cluster is running.

While the emulation of some scenarios and attacks might also be achieved using a simpler topology and with fewer nodes than in the scenario shown in this paper, the capability of the Gotham testbed to emulate complex scenarios that represent real-world network deployments can have multiple benefits. First, the fidelity to emulate complex topologies and node behavior allows the creation of new scenarios that act as digital twins that reflect real network deployments. Organizations can build scenarios to evaluate solutions before using them in production systems, use the testbed as a cyber range and generate relevant network/log datasets for model training instead of solely relying on publicly available datasets to reduce the gap between the experimental and deployment environments. Secondly, using a complex scenario that includes heterogeneous nodes located at multiple network segments opens up the possibility of using the testbed to capture data at different network links for testing new algorithms for distributed computation, such as federated learning~\cite{Konecny2016}, which is gaining relevance to train ML models in distributed IoT scenarios.

Gotham can be extended by increasing the included library of devices, attackers, scenarios, and for sure, using it as a platform to train, implement or validate \textit{superheroes} that react against the attacks from threat actors. We hope that instead of only sharing static datasets for network security that are difficult to adapt for different scenarios and might get outdated, researchers and practitioners can use and build upon the testbed to create and share other complex scenarios for network security that allows the dynamic creation of new datasets tailored to the network setting of interest.



\ifCLASSOPTIONcompsoc
\section*{Acknowledgments}
\else
\section*{Acknowledgment}
\fi

The European commission financially supported this work through Horizon Europe program under the IDUNN project (grant agreement number 101021911). It was also partially supported by the Ayudas Cervera para Centros Tecnológicos grant of the Spanish Centre for the Development of Industrial Technology (CDTI) under the project EGIDA (CER-20191012), and by the Basque Country Government under the ELKARTEK program, project REMEDY - REal tiME control and embeddeD securitY (KK-2021/00091). Urko Zurutuza is part of the Intelligent Systems for Industrial Systems research group of Mondragon Unibertsitatea (IT1676-22), supported by the Department of Education, Universities and Research of the Basque Government.

\bibliographystyle{IEEEtran}
\bibliography{IEEEabrv,referencias,reporeference}


\begin{IEEEbiography}[{\includegraphics[width=1in,height=1.25in,clip,keepaspectratio]{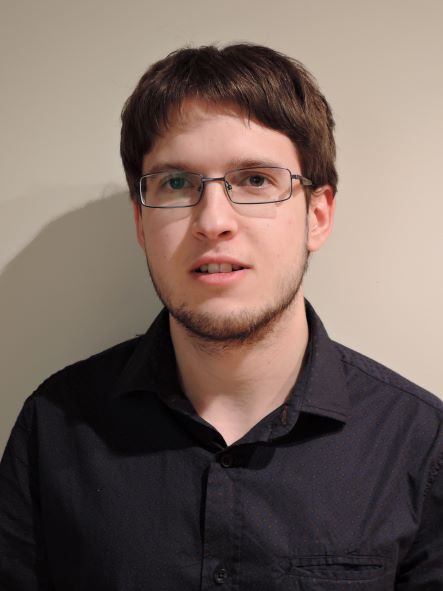}}]{Xabier~Sáez-de-Cámara}
received his B.Sc. degree in Physics and Electronic Engineering from the Faculty of Science and Technology of the Basque Country University in 2015 and 2016, respectively. He holds an M.Sc. in Computational Engineering and Intelligent Systems from the University of the Basque Country. He is currently a Ph.D. student at IKERLAN in the Cybersecurity in Digital Platforms team and the Data Analysis and Cybersecurity research area at Mondragon Unibertsitatea, working on intrusion detection methods in IoT networks.
\end{IEEEbiography}

\begin{IEEEbiography}[{\includegraphics[width=1in,height=1.25in,clip,keepaspectratio]{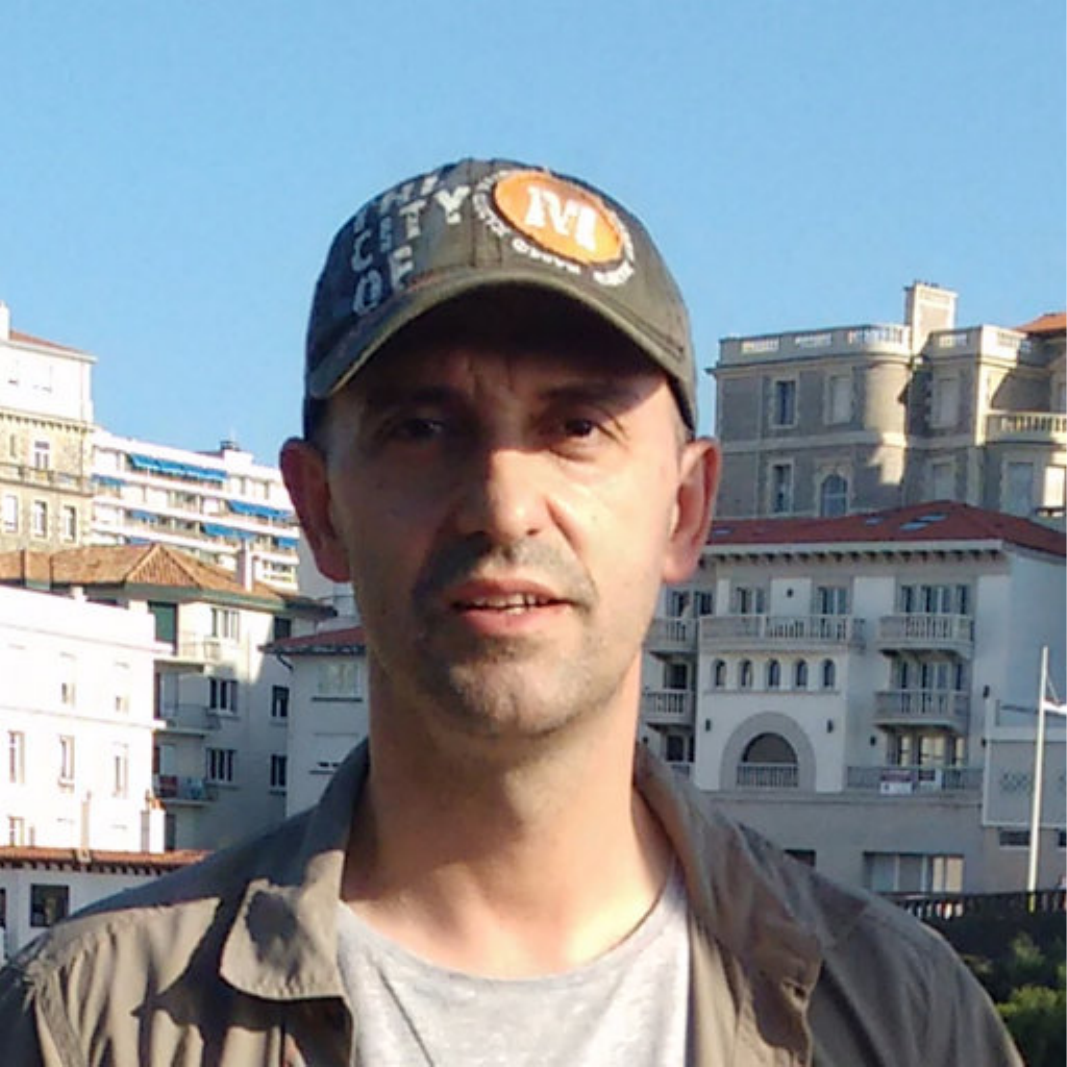}}]{Jose~Luis~Flores} is a researcher at Ikerlan Technology Research Center within the Cybersecurity in Embedded Systems team. He holds a M.Sc.~in Robotics and Advanced Control from the University of the Basque Country. His main interest is related to Artificial Intelligence and Cybersecurity. As such, the main lines he works on in each organization are Embedded System security at Ikerlan, and Machine Learning and Optimization at the university.
\end{IEEEbiography}

\begin{IEEEbiography}
[{\includegraphics[width=1in,height=1.25in,angle=270,clip,keepaspectratio]{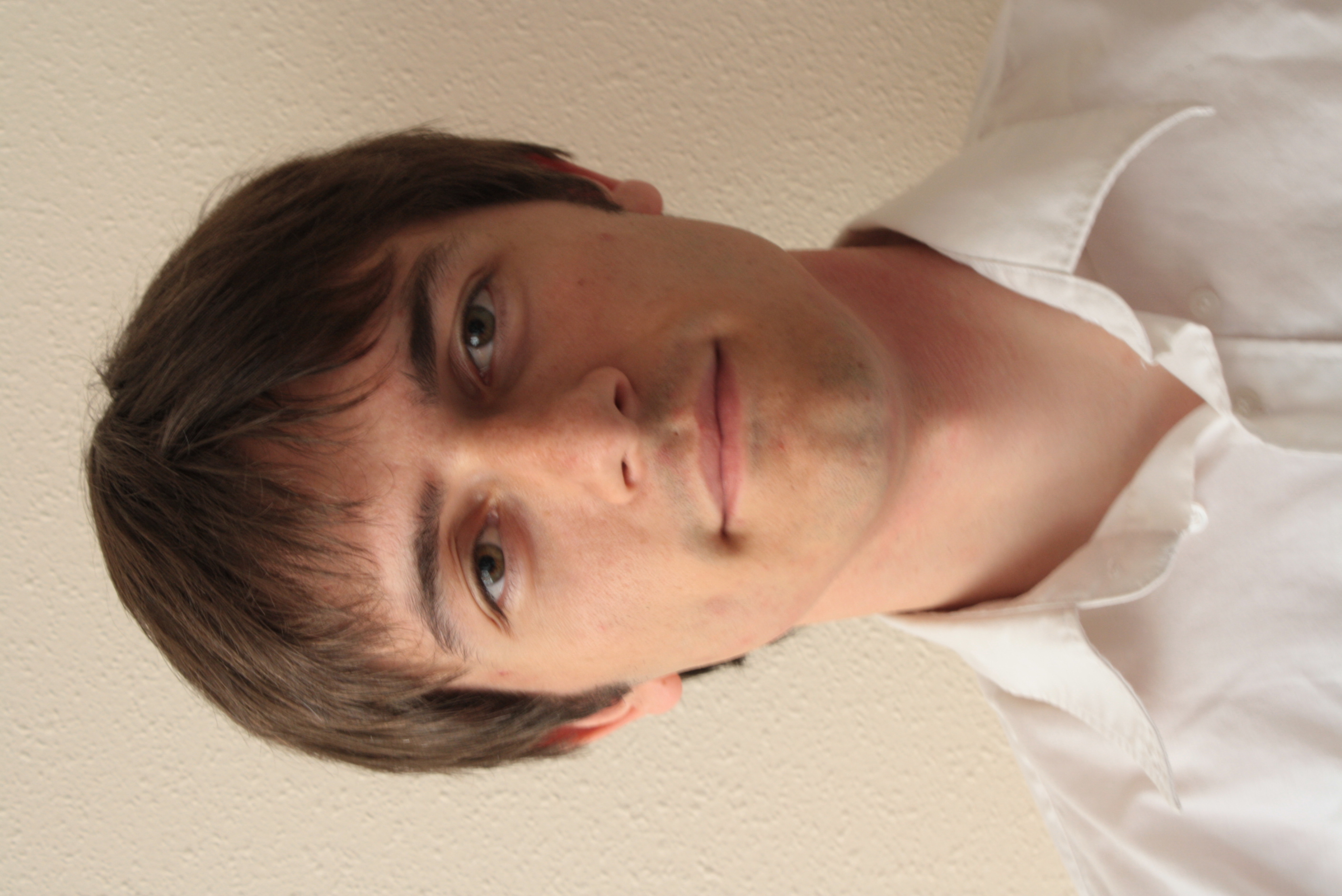}}]{Dr. Cristóbal~Arellano}
studied Computer Engineering at the University of the Basque Country, where he obtained his Ph.D. degree (with international mention) in Web Information Systems in 2013 (Cum Laude unanimously). He has been with IKERLAN since 2015 as a researcher and he currently is part of the Cybersecurity in Digital Platforms team. His current research interests include Cybersecurity in Cloud Platforms, Device Identity Management, DevSecOps, Federated Learning, Vulnerability Monitoring and Threat Detection. He has participated as an author or co-author in conferences such as WWW, ICWE, WISE, etc. He also has participated and led multiple European funded projects such as FP7 MONDO, UTEST, H2020 QUALITY and H2020 IDUNN.
\end{IEEEbiography}

\begin{IEEEbiography}
[{\includegraphics[width=1in,height=1.25in,clip,keepaspectratio]{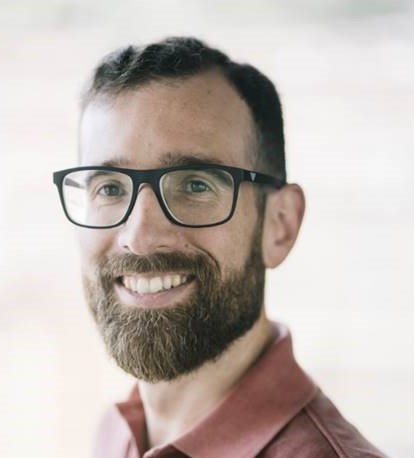}}]{Dr. Aitor Urbieta} studied Computer Engineering at the University of Mondragon, where he obtained his Ph.D. degree (with international mention) in Computer Science in 2010 (Cum Laude unanimously). He has been with IKERLAN since 2007 where he currently leads the Cybersecurity in Digital Platforms research team. His current research interests include Cybersecurity in Digital Platforms, Internet of Things (IoT), Cybersecurity in Communication Protocols, Federated Learning, Blockchain, End-To-End Security, Vulnerability Monitoring, Threat Detection, Fog Computing, Edge Computing and IoT environment validation. He has participated as an author or co-author in more than 30 scientific publications in the previously mentioned areas, some of them Q1, published in national and international conferences and articles in JCR journals.
\end{IEEEbiography}

\begin{IEEEbiography}
[{\includegraphics[width=1in,height=1.25in,clip,keepaspectratio]{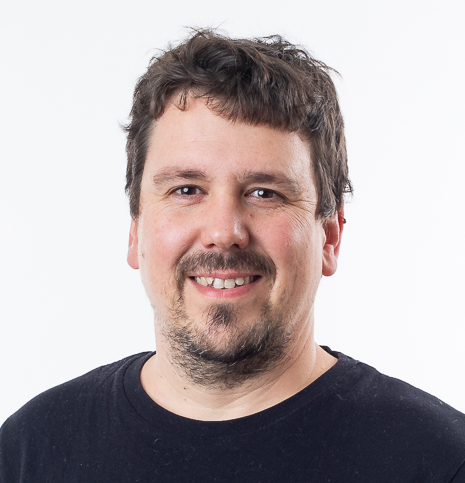}}]{Dr. Urko~Zurutuza}
is the principal investigator of the Intelligent Systems for Industrial Systems research group, and coordinator of the Data Analysis and Cybersecurity research area. He obtained his PhD in January 2008 at Mondragon Unibertsitatea, in collaboration with the Zürich IBM Research Lab. His research interests revolve around applications of Machine Learning to real world problems, and specially Cybersecurity. He has published more than 20 articles in high impact journals, more than 55 publications in blind peer-reviewed conferences, edited 3 books (2 of them as conference proceedings), and coauthored 7 book chapters. He is member of the Board of Directors of RENIC (National Network of Excellence in Cybersecurity Research), and serves in Steering Boards of leading international conferences such as DIMVA or RAID.\end{IEEEbiography}

\end{document}